\documentclass[%
reprint,
amsmath,amssymb,
aps,superscriptaddress
]{revtex4-2}

\usepackage{graphicx}
\usepackage{dcolumn}
\usepackage{bm}

\usepackage{epsfig}
\usepackage[english]{babel}
\usepackage{latexsym}
\usepackage{graphics}
\usepackage{subfigure}
\usepackage{graphicx}
\usepackage{amsmath}
\usepackage{hyperref}
\usepackage{amssymb}
\usepackage{color}

\begin{document}
	
	\preprint{APS/123-QED}
	
	\title{Statistical Characteristics of Tunneling States in Strong-Field Atomic Ionization}

	\author{M. W. Cao}
	\homepage{These authors contribute equally to this paper.}
	\affiliation{College of Physics and Information Technology and Quantum Materials and Devices Key Laboratory of Shaanxi Province's High Education Institution, Shaanxi Normal University, Xi'an, China}
	
	\author{Z. Y. Chen}
	\homepage{These authors contribute equally to this paper.}
	\affiliation{College of Physics and Information Technology and Quantum Materials and Devices Key Laboratory of Shaanxi Province's High Education Institution, Shaanxi Normal University, Xi'an, China}
	
	\author{J. N. Wu}
	\affiliation{College of Physics and Information Technology and Quantum Materials and Devices Key Laboratory of Shaanxi Province's High Education Institution, Shaanxi Normal University, Xi'an, China}

	\author{S. Q. Shen}
	\email{shensqhn@163.com}
	\affiliation{College of Physics and Information Technology and Quantum Materials and Devices Key Laboratory of Shaanxi Province's High Education Institution, Shaanxi Normal University, Xi'an, China}

	\author{S. Wang}
	\email{wangshang@hebtu.edu.cn}
	\affiliation{College of Physics and Hebei Key Laboratory of Photophysics Research and Application, Physics Postdoctoral Research Station, Hebei Normal University, Shijiazhuang, China}
	
	\author{W. Y. Li}
	\email{liwy157@126.com}
	\affiliation{Hebei Key Laboratory of Optoelectronic Information and Geo-detection Technology, School of Mathematics and Science, Hebei GEO University, Shijiazhuang, China}
	
	\author{J. Y. Che}
	\affiliation{College of Physics, Henan Normal University, Xinxiang, China}
	
	\author{Y. J. Chen}
	\email{chenyjhb@gmail.com}
	\affiliation{College of Physics and Information Technology and Quantum Materials and Devices Key Laboratory of Shaanxi Province's High Education Institution, Shaanxi Normal University, Xi'an, China}
	
	\date{\today}
	
	\begin{abstract}
		The state of the tunneling electron under the potential barrier is important in strong laser-atom interaction but is difficult to identify. Recent experiments showed that the tunneling electron may be located in a bound state with high symmetry [Phys. Rev. Lett. 134, 213201 (2025)]. However, the quantitative characteristic of the tunneling state in a tunneling event remains unclear. Here, we study tunneling ionization of atoms in strong circular laser fields. The calculated photoelectron momentum distribution (PMD) through numerical solution of time-dependent Schrödinger equation (TDSE) presents an isotropic ring-shaped distribution and the most probable momentum (MPM) along the ring can be easily identified. The kinetic energy related to MPM is remarkably smaller than that predicted by the strong-field approximation (SFA) that ignores Coulomb potential. Surprisingly, for different target atoms and laser parameters, the kinetic energy difference of MPM between TDSE and SFA is always close to half of the corresponding Coulomb potential at the tunnel exit. This phenomenon can be well described by a proposed model, which indicates that the tunneling electron is in an exit-position-dependent quasibound state agreeing with the virial theorem. 
		These results quantitatively reveal the characteristics of tunneling states from a statistical perspective.   
		
	\end{abstract}
	
	\maketitle
	

	Laser-induced tunneling ionization is at the heart of strong-field physics \cite{Keldysh1965, Faisal1973, Reiss1980}. It can trigger rich strong-field processes such as above-threshold ionization (ATI) \cite{Agostini1979,Ammosov1986,Becker2002}, high-order ATI \cite{Yang1993, Paulus1994, Lewenstein1995}, and high-harmonic generation (HHG) \cite{McPherson1987, Huillier1988, Corkum1993, Lewenstein1994}, etc..   
	Recent studies showed that the ATI or HHG can be used to measure the electron dynamics with unprecedented attosecond time resolution \cite{Eckle2008,Krausz2009,Sha2012,Sainadh2019}. These high-precision measurements require accurate theoretical models to distill attosecond time information from the observables such as PMDs and harmonic spectra. As the ATI and HHG can be well understood by the SFA with the electron-trajectory method \cite{Lewenstein1994,Lewenstein1995,Becker2002}, the important Coulomb potential is neglected in SFA. Many Coulomb-included SFA-based models have been developed to describe the effect of Coulomb potential in recent years, providing deep insights into ATI and HHG mechanisms \cite{Goreslavski2004,Yan2010,Boge2013,Torlina2015,Li2014,Xie2020,Trabert2021}. However, the analytical and accurate description of tunneling electron dynamics under the potential barrier remains a challenging task, as it essentially involves analytical solutions for multi-level atoms in strong laser fields, which are currently difficult to obtain \cite{Scully1997}. Therefore, it is important to study the physical properties of tunneling electron states under the potential barrier through practical and numerical experiments, and to use these properties to construct theoretical models that bypass analytical solutions for multi-level atoms.   
	
	Recently, experimental \cite{Khurelbaatar2025} and theoretical studies \cite{Muller1999,Klaiber2015,Serebryannikov2016,Wu2025} showed that during tunneling, the electron may be located in a highly excited bound state, which plays an important role in the subsequent dynamics of the electron after tunneling. However, the quantitative characteristic of the tunneling state in a specific tunneling event remains unclear. The main difficulty is that the tunneling state is an intermediate state, while the observables are related to the final state of the electron. Therefore, other processes such as the rescattering after tunneling \cite{Yang1993,Corkum1993} can remarkably influence the observables. More specifically, due to that the near-nuclear Coulomb force is much greater than the far-nucleus one, the near-nucleus Coulomb effect dominates in ionization. In the tunneling process, the state of the tunneling electron is closely related to properties of the near-nucleus Coulomb potential. In the rescattering process, when the rescattering electron approaches the nucleus, the near-nucleus Coulomb potential can also play an important role. Consequently, the final observables include not only the near-nucleus Coulomb effect during tunneling but also that during rescattering, making the abstraction of the information for the tunneling state from the observables difficult. The attoclock related to tunneling ionization of atoms and molecules in strong elliptical laser fields with high ellipticity \cite{Eckle2008,Boge2013} provides an exquisite method for studying the tunneling state. In this case, the rescattering electron always moves away from the nucleus. As a result, the observables mainly encode the near-nucleus Coulomb effect in tunneling. However, the typical observable of attoclock, namely the offset angle in PMD, depends on laser ellipticity \cite{Landsman2013,Che2023}, making relevant theoretical studies more complex.  
	
	In this letter, we study the ATI of atoms in strong circularly-polarized laser fields. In this case, the high symmetry of both the laser field and the atomic Coulomb potential results in a simple isotropic ring-shaped distribution of PMD (Fig. 1(a)). Consequently, the weighted mean radius of the circular local most probable route (LMPR) in PMD \cite{Chao2021}, obtained by finding the local maximum in the angular distribution of PMD,  can be used as a characteristic observable to analyze the Coulomb effect instead of the ellipticity-dependent attoclock offset angle. For simplicity, we also call this radius the MPM. 
	
	Our results for He show that the MPM of TDSE increases significantly with increasing laser intensity and wavelength, and under different laser parameters, it is always significantly smaller than that of SFA (Fig. 1(b)). However, surprisingly, for a wide range of laser parameters, the ratio of the kinetic energy difference related to MPM between SFA and TDSE to half of the corresponding Coulomb potential $|V(r_0)|$ at the tunnel exit $\textbf{r}_0$ is always slightly larger than $1.1$.  This ratio is near to or smaller than $1.1$ when a model termed TRCM is used (Fig. 1(d)).  This smaller ratio for TRCM can be attributed to a Coulomb-induced symmetry-related exit velocity $\textbf{v}_c$ considered in the model. Similar phenomena also hold for H and He$^+$ at diverse laser parameters (Fig. 3). These statistical results strongly indicate that the tunneling electron is located in a high-symmetry quasibound state with energy $V(r_0)/2$ near tunnel exit and the exit velocity of the tunneling electron is also rectified by a Coulomb-induced velocity $\textbf{v}_c$ (Fig. 2).

	\textit{Theory methods}. The tunneling electron mainly moves along the laser polarization plane. To simplify our discussions and explore a wide parameter region, our TDSE simulations are first performed for two-dimensional (2D) cases along the polarization plane. Then we extend our considerations to three-dimensional (3D) cases.  The Coulomb potential used here has the form of $V(r)=-Z/\sqrt{x^2+y^2+\xi}$ for 2D cases and $V(r)=-Z/\sqrt{x^2+y^2+z^2+\xi}$ for 3D cases. Here, $\xi$ is the smoothing parameter with $\xi=0.5$ for 2D cases and $\xi=0.071$ for 3D cases, and $Z$ is the effective charge which is adjusted so that the ground-state ionization potential $I_p$ of the atom reproduced here matches the actual one. We choose the target atoms of He, H and He$^+$ having different ionization potentials of $I_p=0.9$ a.u., $0.5$ a.u. and $2$ a.u.. The TDSE of $i\frac{\partial}{\partial t}\Psi(\mathbf{r},t)= H(t)\Psi(\mathbf{r},t)$ is solved numerically using the spectral method \cite{Feit1982}.  We use a grid size of $L_x\times L_y(\times L_z) = 409.6 \times 409.6 (\times 51.2)$ a.u. for 2D (3D) simulations and the space step used is $0.4$ a.u. for each dimension. The time step used is $\Delta t= 0.05$ a.u. The numerical convergence is checked by using smaller time steps and finer grids. More details for TDSE calculations can be found in \cite{Che2023}.

	To extract the information of the Coulomb effect from the calculated MPM $p$ of TDSE, we compare the TDSE results of $p_{TDSE}$ with the SFA results of $p_{SFA}$ where the Coulomb potential $V(r)$ is neglected and with results $p_{TRCM}$ of a semiclassical SFA-based model, which is called the tunneling-response-classic-motion (TRCM) model \cite{Che2021,Che2023,CheJ2023}. The TRCM assumes that due to the quantum property of the near-nucleus Coulomb effect, the tunneling electron at the tunnel exit $\textbf{r}_0$ is located in a quasibound state with the average energy of $\langle\textbf{v}^2/2\rangle+\langle{V(r)}\rangle\approx V(r_0)/2$, which agrees with the virial theorem of  $\langle\textbf{v}^2/2\rangle=n_f\textbf{v}_c^2/2\approx -\langle{V(r)}\rangle/2=-V(r_0)/2$. Here, the parameter $n_f=2$ or $3$ is the dimension of the system studied and reflects the space symmetry of the system \cite{Guo2025}. The velocity $\textbf{v}_c$ will be introduced later. Beyond the exit position $\textbf{r}_0$, the Coulomb effect can be neglected. In addition, the TRCM also assumes that the near-nucleus Coulomb effect will induce a velocity $\textbf{v}_c=-v_c\textbf{r}_0/r_0$ at the tunnel exit, which is antiparallel to the exit position $\textbf{r}_0$ \cite{Goreslavski2004}. The amplitude of $\textbf{v}_c$ is given by the root mean square velocity ${v}_c\approx\sqrt{\left| V({r}_0) \right|/n_f}$ relating to the virial theorem. Due to this velocity, the MPM in TRCM satisfies ${p}_{TRCM}=|\textbf{p}_{SFA}+\textbf{v}_c|$. It has been shown that the velocity $\textbf{v}_c$ plays a critical role in the attoclock offset angle \cite{Che2023}. The calculation details for SFA and TRCM can also be seen in \cite{Che2023}. In short, in SFA, for each drift momentum $\textbf{p}$, we solve the saddle-point equation $[\textbf{p}+\textbf{A}(t)]^2/2=-I_p$ to obtain the complex saddle-point time $t_s=t_0+it_x$. Here, $\textbf{A}(t)$ is the vector potential of the laser electric field $\textbf{E}(t)$. Then we can obtain the amplitude $c(\textbf{p})\propto e^{b}$ for the momentum $\textbf{p}$. Here, $b$ is the imaginary part of the quasiclassical action $S(\textbf{p},t_s)=\int_{t_s}\lbrace[\textbf{p}+\textbf{A}(t')]^2/2+I_p\rbrace dt'$. In TRCM, the Coulomb-modified drift momentum $\textbf{p}_c$ satisfies the relation of $\textbf{p}_c=\textbf{p}+\textbf{v}_c$.  The amplitude for $\textbf{p}_c$ can be written as $c'(\textbf{p}_c)\equiv c(\textbf{p})$. Then we can obtain the PMDs $|c(\textbf{p})|^2$ ($|c'(\textbf{p}_c)|^2$) of SFA (TRCM). The exit position $\textbf{r}_0$, which is used in evaluating $\textbf{v}_c$ and differs for different tunneling events, can be calculated through the expression of $\mathbf{r}_0\equiv \mathbf{r}\left( t_0 \right) =Re \left( \int_{t_{s}}^{t_0}{\left[\mathbf{p}+\mathbf{A}\left( t' \right) \right] dt'} \right)$ \cite{Yan2010}. We consider a wide range of laser parameters with Keldysh parameters \cite{Keldysh1965} being smaller than $1$ at which SFA and TRCM work better \cite{Chen2025}. 
	
	\begin{figure}[t]
		\begin{center}
			\rotatebox{0}{\resizebox *{8.5cm}{8cm} {\includegraphics {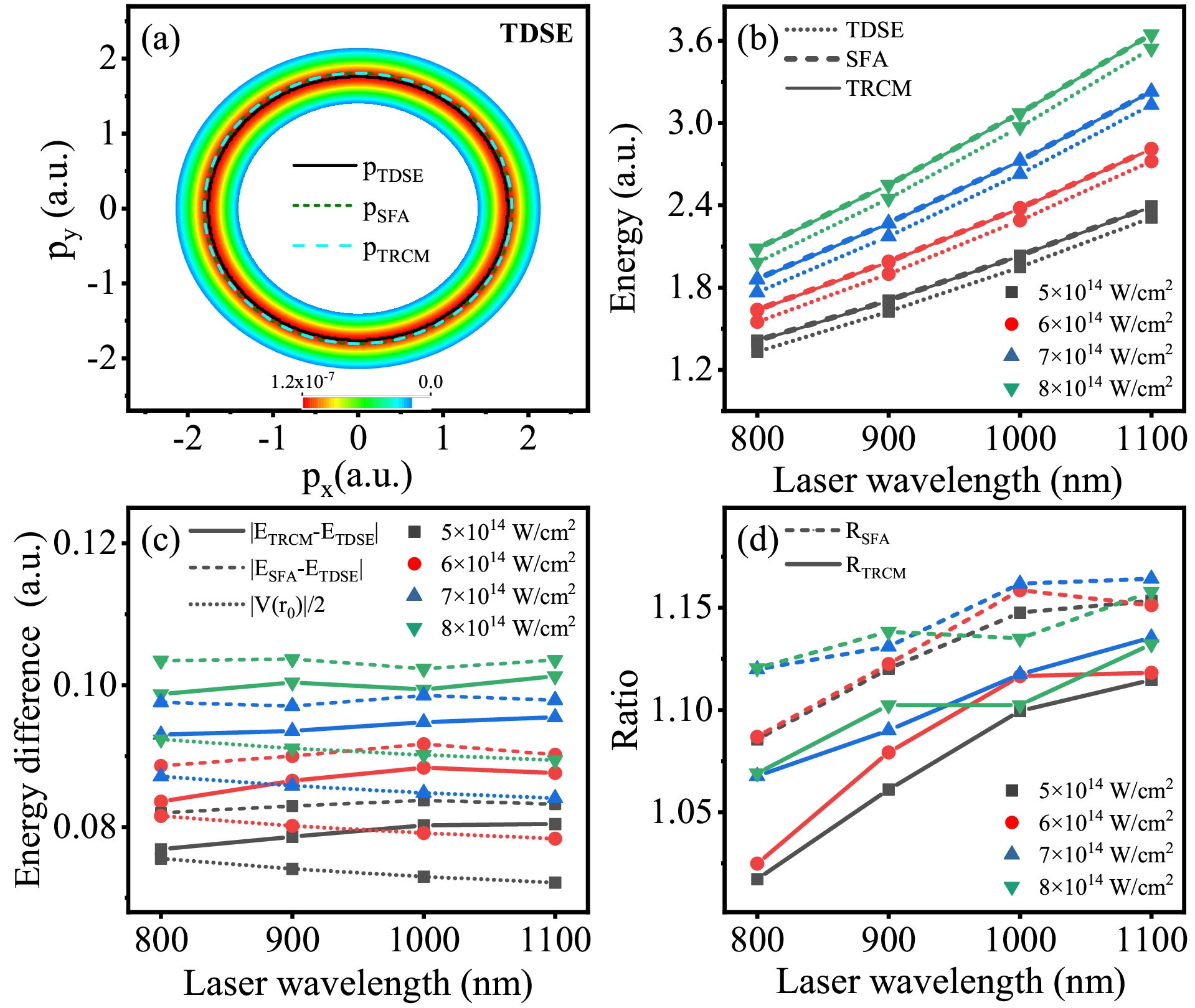}}}
		\end{center}        
		\caption{Results for ionization of 2D He obtained by different methods in circular laser fields with different laser parameters. In (a), we show the PMD of TDSE at $I=6\times10^{14}$W/cm$^2$ and $\lambda=800$ nm. The LMPRs with the radii (MPM)  $p_{TDSE(SFA,TRCM)}$, obtained through TDSE, SFA and TRCM, respectively, are also shown here. In (b), we show the most probable kinetic energy $E_{TDSE(SFA,TRCM)}=p_{TDSE(SFA,TRCM)}^2/2$. In (c), we show the energy difference of $E_{SFA(TRCM)}-E_{TDSE}$. The corresponding values of  $\left|V({r}_0)\right|/2$ are also shown here. In (d), we show the ratio of $R_{SFA(TRCM)}=(E_{SFA(TRCM)}-E_{TDSE})/(\left|V({r}_0)\right|/2)$.  Laser parameters used in (b)-(d) are as shown.}
		\label{fig:g1}
	\end{figure}
	
	\textit{Typical characteristics}. In Fig. 1(a), we show the PMD of He obtained with 2D TDSE. The PMD shows an isotropic ring-shaped distribution, and the LMPR related to the local maximal amplitude in this distribution \cite{Chao2021} presents a circular structure, as shown by the black-solid line. For comparison, we also present the LMPRs of SFA and TRCM. The radii of these circular LMPRs give the corresponding MPMs. In Fig. 1(b), we show the most probable kinetic energy $E_p$ related to the MPM ${p}$ obtained using different methods. This energy is higher for higher laser intensities and longer wavelengths, which can be easily understood by the classical model of $\textbf{p}\approx -\textbf{A}(t)$  \cite{Corkum1993}. In particular, for specific laser parameters, this energy of SFA is slightly greater than that of TRCM, but significantly larger than that of TDSE, with a difference around $0.09$ a.u.. This difference is highlighted in Fig. 1(c). For a fixed laser wavelength, this difference increases with the increase of laser intensity, while for a fixed laser intensity, this difference first increases when increasing the wavelength, then becomes insensitive to the wavelength.  In Fig. 1(c), we also show the value of $|V(r_0)|/2$. 
	For different laser parameters, this value is slightly smaller than the energy difference between TRCM (SFA) and TDSE. In particular, as the laser intensity increases, the curve of $|V(r_0)|/2$ becomes approximately parallel to the energy-difference curve. 
	
	Further comparisons are plotted in Fig. 1(d), where we show the ratio of the energy difference between SFA (TRCM) and TDSE to $|V(r_0)|/2$. One can observe that this ratio related to SFA is around $1.125$ for varied laser parameters, and it is near to or smaller than $1.1$ for that related to TRCM overall. Since the main difference between TRCM and SFA is related to the Coulomb-induced velocity $v_c\approx\sqrt{|V(r_0)|/2}$, this ratio difference between TRCM and SFA can be attributed to the velocity $v_c$. The applicability of the expression for $v_c$ has been validated by comparing the predictions of TRCM to experiments for attoclock offset angles of different target atoms and laser parameters \cite{Che2021,Che2023,CheJ2023}.

	\begin{figure}[t]
		\begin{center}
			\rotatebox{0}{\resizebox *{8.5cm}{5.3cm} {\includegraphics {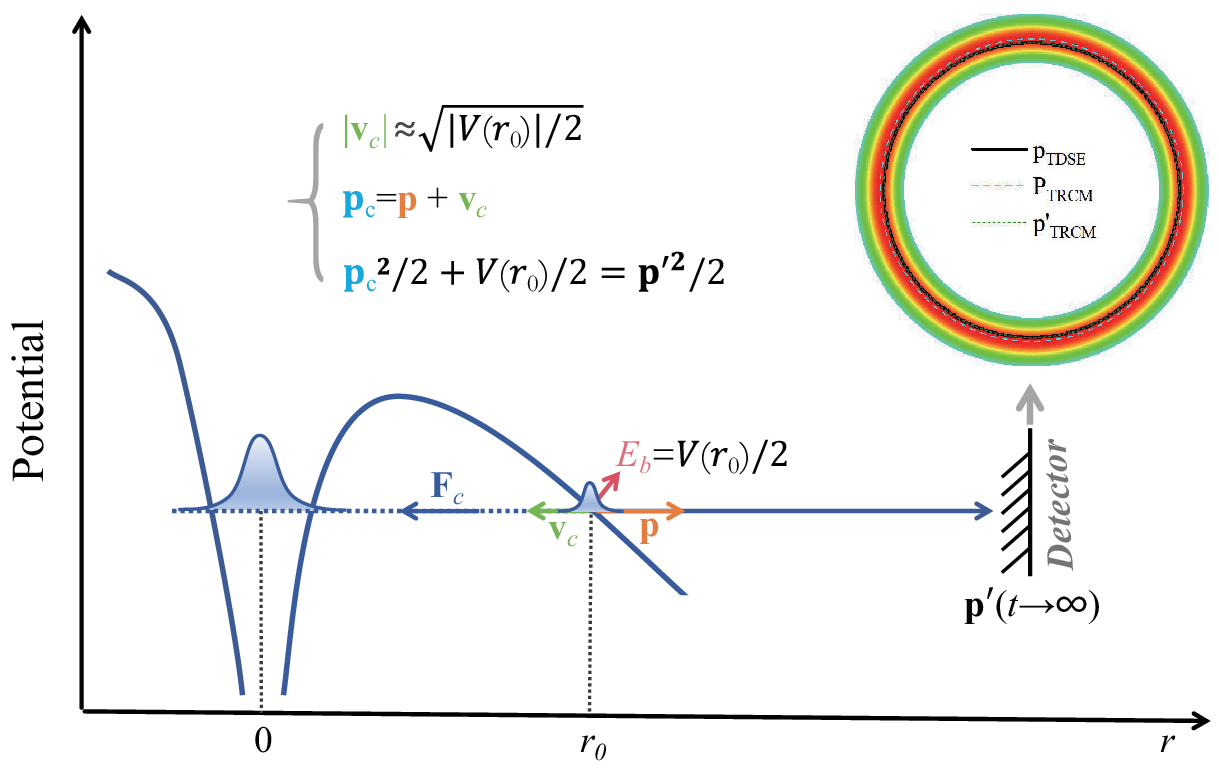}}}
		\end{center}
		
		\caption{Sketch of laser-induced tunneling ionization of atoms. The Coulomb potential of the atom is bent by the laser field, forming a barrier (blue-solid line) through which the bound electron (the wave packet at $\textbf{r}=0$) can tunnel. Without considering the Coulomb, according to SFA, the electron born at the tunneling-out time $t_0$ will appear at the exit position $\textbf{r}_0$ with the exit velocity $\textbf{v}(t_0)$ and has the drift momentum $\textbf{p}=\textbf{v}(t_0)-\textbf{A}(t_0)$. When considering the Coulomb, 1) during the process of the tunneling electron moving from the origin towards $\textbf{r}_0$, it will be subject to the Coulomb force $\textbf{F}_c$. From a semiclassical perspective, this force will induce a velocity $\textbf{v}_c$ at the tunnel exit, whose direction is opposite to the exit position $\textbf{r}_0$. According to TRCM, $v_c\approx\sqrt{\left| V({r}_0) \right|/n_f}$ with $n_f=2$ for 2D cases.  Accordingly, the Coulomb-corrected drift momentum of TRCM is $\textbf{p}_{c}=\textbf{p}+\textbf{v}_c$. 2) In addition, according to TRCM, the tunneling electron will be located in a quasibound state (the wave packet at $\textbf{r}=\textbf{r}_0$),   which agrees with the virial theorem and has the average energy of $V(r_0)/2$.  As a result, the total energy $E_{total}$ of the tunneling electron is $E_{total}=\textbf{p}'^2/2=\textbf{p}_{c}^2/2+V(r_0)/2$. Here, $\textbf{p}' $ is the prediction of the developed TRCM for the final observable momentum at the detector, which also considers the energy correction $V(r_0)/2$. The color coding shows the PMD obtained using TRCM for 2D He. For comparison, the LMPRs of TRCM and TDSE with the radii (MPM) $p_{TDSE}$ (black-solid circle) and $p_{TRCM}$ (blue-dashed circle), and the route of the developed TRCM with the radius $p'_{TRCM}$ (green-dotted circle) are also shown here. }
		\label{fig:g2}  
	\end{figure}
	
	\textit{Implications on tunneling}. The tunneling mechanism implied by the results in Fig. 1 is plotted in Fig. 2.  Firstly, we discuss the Coulomb effect on the exit velocity of the tunneling electron. Without considering Coulomb, at the tunnel exit $\textbf{r}_0$ at the tunneling-out time $t_0$, according to the SFA, the electron will have the exit velocity $\textbf{v}(t_0)$ corresponding to the drift momentum $\textbf{p}=\textbf{v}(t_0)-\textbf{A}(t_0)$ \cite{Che2023}. For simplicity, we only indicate the drift momentum $\textbf{p}$ at the tunnel exit in Fig. 2. However, when the Coulomb potential is considered, the Coulomb force $\textbf{F}_c$ will pull the electron back when it tunnels through the barrier under the action of the laser electric field. This will equivalently induce an additional velocity $\textbf{v}_c$ at the tunnel exit, with the direction opposite to the exit position $\textbf{r}_0$. The value of $v_c$ can not be directly evaluated through classical methods, as the quantum effect dominates near the nucleus. It can be approximately evaluated by the root mean square velocity $v_c$ of TRCM \cite{Chen2025}. Accordingly, the Coulomb-corrected drift momentum $\textbf{p}_c$ can be written as $\textbf{p}_c=\textbf{p}+\textbf{v}_c$. Secondly, we discuss the Coulomb effect on the energy of the tunneling electron. Due to the Coulomb potential, at the tunnel exit, the tunneling electron is located in a quasibound state with energy $V(r_0)/2$. Therefore, the total energy of the tunneling electron including laser-dominating kinetic energy and Coulomb-related potential energy is $E_{total}=\textbf{p}_c^2/2+V(r_0)/2$. When the tunneling electron moves from the tunnel exit to the detector, the negative potential energy $V(r_0)/2$ will be offset by the kinetic energy $\textbf{p}_c^2/2$, resulting in the final observable momentum $p'=\sqrt{2E_{total}}$. 
	
	The inset in Fig. 2 shows the PMD of TRCM for 2D He. The general TRCM only considers the Coulomb-related velocity correction $\textbf{v}_c$, and predicts the Coulomb-modified drift momentum $\textbf{p}_c=\textbf{p}+\textbf{v}_c$ \cite{Che2023}. This gives the MPM $p_{TRCM}=\sqrt{\textbf{p}_c^2}$ for the momentum $\textbf{p}_c$ having the maximal amplitude. The value of $p_{TRCM}$ predicted by the general TRCM differs remarkably from the TDSE result $p_{TDSE}$. To further include the Coulomb-related energy correction $V(r_0)/2$ into TRCM,  the MPM obtained by the developed TRCM is $p'_{TRCM}=\sqrt{\textbf{p}_c^2+V(r_0)}$, which agrees with $p_{TDSE}$, as seen in the inset.

	\begin{figure}[t]
		\begin{center}
			\rotatebox{0}{\resizebox *{8.5cm}{6.5cm} {\includegraphics {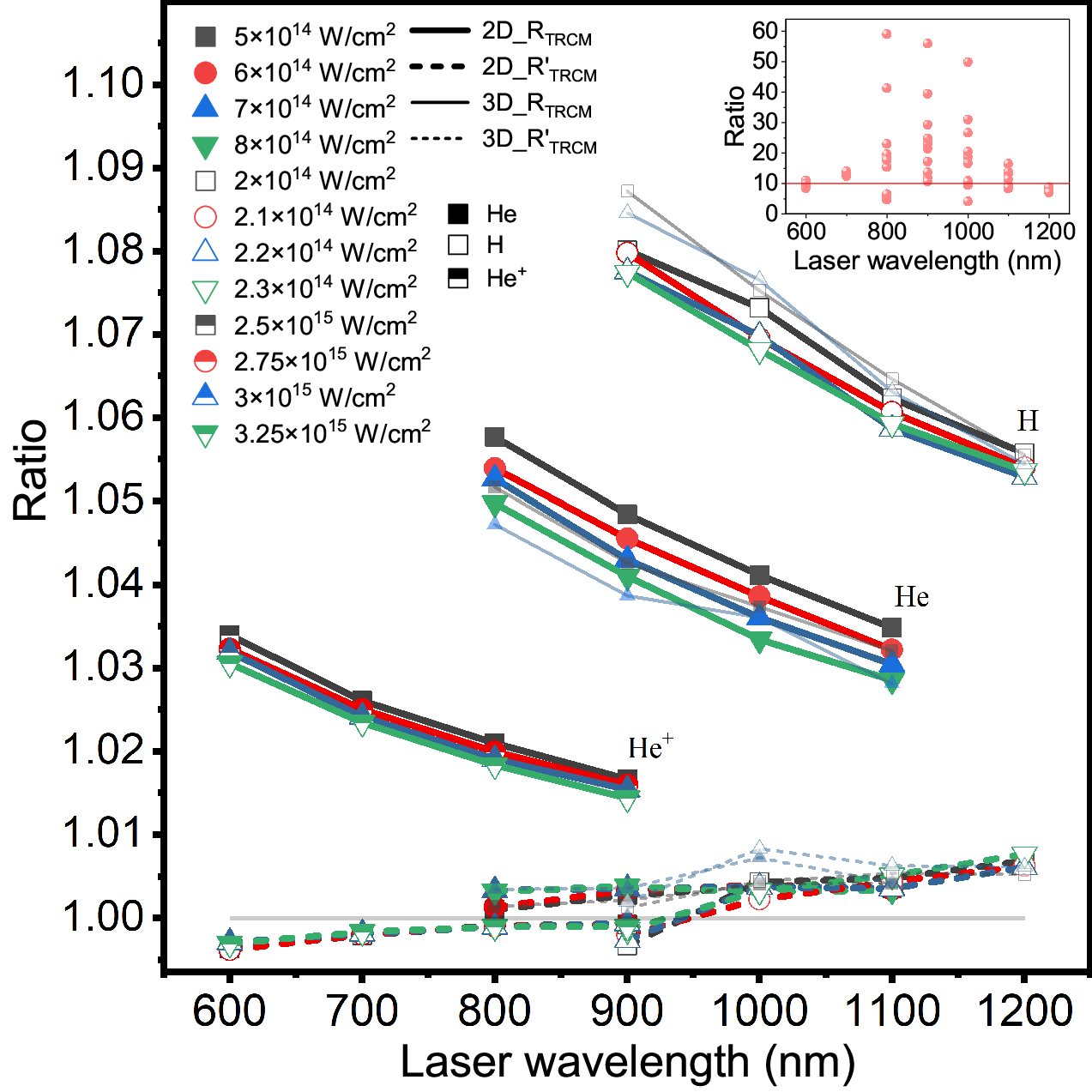}}}
		\end{center}
		
		\caption{Ratios of $R_{TRCM}=(p_{TRCM}^2/2)/(p_{TDSE}^2/2)$ (solid lines) and $R'_{TRCM}=(p_{TRCM}^2/2+V(r_0)/2)/(p_{TDSE}^2/2)$ (dashed lines) for different targets of 2D (bold lines) and 3D (thin lines) He (solid symbols), H (hollow symbols) and He$^+$ (half-hollow symbols) at different laser intensities and wavelengths as shown. The light-gray line at the unity is plotted to guide the eyes. The inset shows the ratios of $(R_{TRCM}-1)/(R'_{TRCM}-1)$ for these 2D and 3D results here.}
		\label{fig:g3}
	\end{figure}
	
	\textit{Cases of different targets}. To validate the above discussions, we further perform simulations for different targets and relevant results are shown in Fig. 3. Here, we compare the two ratios of $R_{TRCM}=(p_{TRCM}^2/2)/(p_{TDSE}^2/2)$ and $R'_{TRCM}=(p_{TRCM}^2/2+V(r_0)/2)/(p_{TDSE}^2/2)$ for 2D He, H and He$^+$ at varied laser parameters. When the ratios of $R_{TRCM}$, which do not consider the Coulomb effect on the energy, fall within the range of $1.015$ for He$^+$ at high laser intensities to $1.08$ for H at relatively low laser intensities, the corresponding ratios of $R'_{TRCM}$ fall within the range of $0.995$ to $1.007$, very near to $1$. We have also performed simulations for 3D He and H and relevant results are also shown in Fig. 3, which are similar to 2D ones. The 2D and 3D data of $R_{TRCM}$ and $R'_{TRCM}$ in Fig. 3 indicate that the consideration of the energy correction $V(r_0)/2$ greatly improves the agreement between TDSE and TRCM. 
	The remaining difference (i.e., the small deviation of the ratio $R'_{TRCM}$ from $1$) may come from the approximate description of saddle-point method for ionization \cite{Chen2025}.
	In fact, our further analyses show that the ratios of $|R_{TRCM}-1|/|R'_{TRCM}-1|$ (solid spheres) for these data are larger than $10$ (the horizontal line) on the whole, as shown in the inset.  These results strongly support the energy relation $E_{total}=\textbf{p}_{TDSE}^2/2\approx\textbf{p}_{TRCM}^2/2+V(r_0)/2$, which suggests that at the tunnel exit $\textbf{r}_0$, the tunneling electron is in a quasibound state that has the average energy of $V(r_0)/2$. This is different from the classical case, where the electron appearing in the potential $V(r)$ has an energy of $V(r_0)$ at position $\textbf{r}_0$.  
	
	\textit{Conclusion}. In summary, we have studied tunneling ionization of atoms in strong circular laser fields. The calculated PMD shows an isotropic ring-shaped distribution and the most probable momentum (MPM) along the ring is easy to determine. This enables us to  identify the state of the tunneling electron near the tunnel exit $\textbf{r}_0$ by analyzing  the MPM. The kinetic energy related to MPM obtained through TDSE is smaller than that evaluated using SFA, and the difference in energy between TDSE and SFA is near $V(r_0)/2$, in agreement with the prediction of a Coulomb-included model called TRCM. The TRCM predicts at the tunnel exit, the tunneling electron is located in a high-symmetry quasibound state with the average energy $V(r_0)/2$. In addition, the near-nucleus Coulomb potential also induces a velocity $\textbf{v}_c$ with $v_c\approx\sqrt{|V(r_0)|/n_f}$ at the tunnel exit, which is antiparallel to the exit position $\textbf{r}_0$ and prevents the tunneling electron from escaping. In practice, the general TRCM only considers the effect of $\textbf{v}_c$ on the observables. With further considering the effect of $V(r_0)/2$, the kinetic energy of MPM predicted by the developed TRCM is always very near to the TDSE results for different targets and laser parameters. Our statistical findings in the work provide direct evidence for the quantitative and semiclassical properties of tunneling states, which are of fundamental importance for understanding tunneling mechanisms and constructing accurate models of strong-field tunneling ionization of atoms and molecules.

	\textit{Acknowledgements}. This work was supported by the National Natural Science Foundation of China (Grant Nos. 12574376, 12547162, 12404330, 12304303).  

\end{document}